\documentclass[twocolumn,showpacs,preprintnumbers,amsmath,amssymb,prb,aps]{revtex4-2} 
\usepackage{epsfig}
\usepackage{epstopdf}
\usepackage{multirow}
\usepackage{graphicx}
\usepackage{dcolumn}
\usepackage{bm}
\usepackage{textcomp}
\usepackage{array}
\usepackage{csquotes}
\usepackage{subcaption}
\usepackage{booktabs}
\usepackage{amsmath}
\usepackage{hyperref}
\hypersetup{
    colorlinks=true,
    linkcolor=blue,
    filecolor=magenta,      
    urlcolor=red,
    pdftitle={},
    }
    
\begin{document}
\title{Realization of strain induced multiple topological phases \\in Cu$_2$SnS$_3$: An \textit{ab-initio} study}
\author{Prakash Pandey$^{1}$}
\altaffiliation{ \url{prakashpandey6215@gmail.com}}
\author{Sudhir K. Pandey$^{2}$}
\altaffiliation{ \url{sudhir@iitmandi.ac.in}}
\affiliation{$^{1}$School of Physical Sciences, Indian Institute of Technology Mandi, Kamand - 175075, India\\$^{2}$School of Mechanical and Materials Engineering, Indian Institute of Technology Mandi, Kamand - 175075, India}

\date{\today}

\begin{abstract}

The search of multiple topological phases (TPs) and their transitions by tuning different parameters through chemical substitutions, electric field, magnetic field, strain and Floquet engineering, etc has garnered a widespread attention in recent time. In spite of great effort, the observations of multiple TPs in a single material and multiple TP transitions in the presence of one parameter remain elusive. Here we demonstrate the presence of multiple TPs and their transitions with uniaxial compressive strain (UCS) in orthorhombic Cu$_2$SnS$_3$ by using \textit{state-of-the-art} \textit{ab-initio} calculations. In the absence of spin-orbit coupling (SOC), the Cu$_2$SnS$_3$ exhibits only one (type-II) nodal-ring and in the presence of SOC, it hosts Weyl phase with seven Weyl points (three at $\Gamma$ and four at general positions) along with nodal arcs. 
On the application of UCS, the system exhibits a type-II nodal ring for UCS $<5.5$\%, which further evolves into type-III nodal-ring for $5.5\% \leq$ UCS $<5.6$\%, thereby marking a TP transition in the nodal topology.
Interestingly, at 5.6\% of UCS, it shows Weyl phase with four Weyl nodes even in the absence of SOC. All the above-mentioned seven Weyl points persist below $5$\% of UCS. For 5\% $\leq$ UCS $<5.6$\%, four Weyl points (at general positions) disappear and nodal-arcs remain intact in all the studied range of UCS. The TPs observed in the absence of SOC appears to arise due to the presence of strain driven topological flat band, which is typically reported to be seen in kagome and Lieb lattices.

\end{abstract}

\maketitle

{\it Introduction.|}
The emergence of exotic fermionic quasiparticles has fundamentally transformed our understanding of quantum states over the past few decades~\cite{PhysRevLett.45.494, PhysRevLett.50.1395}.
Dirac fermions in graphene~\cite{RevModPhys.81.109, novoselov2005two}, helical Dirac fermions at the surface of three-dimensional topological insulators~\cite{RevModPhys.82.3045, 10.7566/JPSJ.82.102001}, Majorana fermions in superconducting heterostructures~\cite{10.1126/science.1222360, 10.1126/science.1259327}, and the discovery of Weyl/Dirac semimetals~\cite{RevModPhys.90.015001, 10.1126/sciadv.1603266} recently mark new frontiers in this field. These topological nontrivial states are driven by different fermionic quasiparticle excitations~\cite{PhysRevLett.119.206402}, including spin-1 fermions with threefold degeneracy carrying $\pm2$ topological charge, spin-3/2 Rarita-Schwinger-Weyl fermions with fourfold degeneracy carrying $\pm4$ topological charge~\cite{10.1126/science.aaf5037, PhysRevLett.119.206402}, double Weyl fermions with $\pm2$ topological charge~\cite{PhysRevLett.108.266802}, and double spin-1 fermions with sixfold degeneracy~\cite{10.1126/science.aaf5037}.


Usually, the energy spectrum of these quasiparticles excitations are described by $E_{\pm} (\textbf{k})=\sum_i k_iA_{i0} \pm \sqrt{\sum_j (\sum_i k_i A_{ij})^2}=T(\textbf{k})\pm U(\textbf{k})$~\cite{soluyanov2015type}. Here, $T(\textbf{k})$ and $U(\textbf{k})$ are the kinetic and potential components of the energy spectrum. The most general Hamiltonian describing a Weyl point is $H (\textbf{k})=\sum_{ij} k_i A_{ij}\sigma_j$~\cite{soluyanov2015type}, where $k_i$ are components of the wave vector \textbf{k}, $\sigma_j$ are the Pauli matrices, and $A_{ij}$ represents the coupling coefficients between the $k_i$ and $\sigma_j$.  
The term $T(\textbf{k})$, which is linear in momentum, decides the tilting of the cone-like spectrum. For example, if there exists a direction \textbf{k} for which $T (\textbf{k}) > U (\textbf{k})$, then a type-II touching point appears~\cite{soluyanov2015type, 10.1126/sciadv.1603266, He_2018}; otherwise, band touching is type-I. Also, the linear dispersion along a specific \textbf{k} direction is classified based on the behavior of the two touching bands: (i) opposite slopes correspond to type-I; (ii) identical slopes define type-II~\cite{soluyanov2015type, PhysRevB.96.081106, Liu_2024}; and (iii) one linear band with zero slope while that of the other one is nonzero is termed type-III nodal line~\cite{PhysRevB.103.L081402, PhysRevB.106.195429, Liu_2024}.
In addition to this, the classification of different types is based not only on the typical conical dispersion but also on the topology of the constant energy surface (referred to as the Fermi surface when the node point lies at the Fermi level). For example, the type-I Weyl phase has both a conical dispersion and a point-like constant energy surface, whereas the type-II phase exhibits an overtilted cone-shaped dispersion, with both electron and hole pockets that touch at the nodal point~\cite{soluyanov2015type, PhysRevB.103.L081402}. The type-III Weyl phase is also a protected band crossing point, but it appears at the contact point of a line-like constant energy surface, which makes it distinct from both type-I and type-II~\cite{PhysRevB.103.L081402, nissinen2017type}. Moreover, type-I, type-II, and type-III Weyl semimetal phases are also distinguished based on their characteristic density of states near the Fermi level, which are vanishing, exhibiting a parabolic peak, and finite, respectively~\cite{PhysRevB.98.121110}. It is typically seen that different types of Weyl phases exhibit distinct manifestations of the chiral anomaly: in type-I, it appears along all directions; in type-II, it is anisotropic; and in type-III, it is absent except within the critical plane~\cite{PhysRevLett.117.077202, PhysRevLett.117.086401}. These types of topological phases not only exhibit novel physical properties but also provide a versatile platform for simulating relativistic particles from high-energy physics. For example, the type-II and type-III semimetallic phases have been theoretically proposed as solid-state analogs corresponding to the inside and the horizon of a black hole, respectively~\cite{volovik2016black, volovik2018exotic, PhysRevB.98.121110}. These phases are crucial and are likely to open up new directions for studying cosmological events in the lab~\cite{PhysRevX.7.041026, PhysRevLett.46.1351, PhysRevLett.85.4643}.

To realize the type-III nodal ring, it is necessary to identify the topological flat band (TFB) on a specific plane in \textbf{k}-space. 
A TFB refers to dispersionless energy bands spanning the entire Brillouin zone, believed to be crucial for realizing fractional topological states without Landau levels~\cite{PhysRevLett.106.236802, PhysRevLett.106.236803, PhysRevLett.107.146803}. 
The interplay between nontrivial topology and flat bands is noteworthy, leading to the formation of nontrivial TFBs that serve as an ideal platform for realizing various topological phases associated with flat bands, including type-III quasiparticles such as Weyl, Dirac, and nodal line phases. There are various lattices that give rise to a TFB, such as the kagome~\cite{PhysRevLett.106.236802, Mielke_1992}, Lieb~\cite{PhysRevLett.62.1201, 10.1143/PTP.99.489}, dice~\cite{PhysRevB.84.241103}, pyrochlore~\cite{PhysRevLett.103.206805}, and perovskite lattices~\cite{PhysRevB.82.085310}. 
Starting with Lieb's work on the Hubbard model~\cite{PhysRevLett.62.1201} and subsequently extended by Mielke~\cite{Mielke_1991_1} and Tasaki~\cite{PhysRevLett.69.1608}, research has expanded to include various lattice models based on line-graph construction~\cite{Mielke_1991_1, Mielke_1991_2}, such as the kagome~\cite{PhysRevLett.106.236802, Mielke_1992}, Lieb~\cite{PhysRevLett.62.1201, 10.1143/PTP.99.489}, breathing-kagome~\cite{PhysRevLett.120.026801, PhysRevB.99.165141}, diatomic-kagome~\cite{PhysRevB.102.125115}, coloring-triangle~\cite{PhysRevB.99.100404}, and diamond-octagon lattices~\cite{PhysRevB.98.245116}, as well as more recent studies on square and honeycomb lattices~\cite{PhysRevLett.99.070401, PhysRevLett.113.236802}. Researchers generally use these lattices to design type-III nodal rings, but promising topological phases associated with TFBs can also be achieved using external parameters, such as strain~\cite{PhysRevB.109.035167}, in simple lattices~\cite{PhysRevX.11.031017}. The aforementioned lattices have established material physics as a promising platform for realizing nontrivial physical phenomena, such as the excitonic insulator state~\cite{PhysRevLett.126.196403} and the excited quantum anomalous/spin Hall effect~\cite{zhou2022excited}.


However, apart from these, only a few electronic TFB materials have been predicted, despite extensive theoretical and experimental studies~\cite{PhysRevB.94.081102, PhysRevB.93.165401, PhysRevB.101.161114, Silveira_2017, gao2020quantum}.
Islam \textit{et al.}~\cite{PhysRevResearch.4.023114} recently studied three-dimensional TFB in superlattices of the HgTe class of materials. In their work, they reported a nodal line semimetal characterized by two isoenergetic, circular nodal-rings at the Fermi level. Furthermore, under compressive hydrostatic pressure, they found that the superlattice exhibits a trivial insulating state, whereas the unstrained superlattice behaves as an ideal Weyl semimetal. However, in this class of superlattice materials, topological phases evolve in the presence of hydrostatic pressure. Naturally, two unresolved questions arise: First, can the coexistence of a TFB and nodal line near the Fermi level be realized without requiring superlattice? Second, can these features be tailored by applying external strain? To our knowledge, the coexisting features has not been explored in orthorhombic body-centered lattices. 
However, multiple topological phases and their transitions have been observed to arise by tuning different parameters through chemical substitutions, electric fields, magnetic fields, strain, and Floquet engineering, etc~\cite{yuan2013zeeman, PhysRevB.80.121308, PhysRevX.9.021028, PhysRevB.108.235166, PhysRevB.111.045406}.
In the present work, we systematically investigate the coexistence of TFB with nodal line semimetals in Cu$_2$SnS$_3$ using \textit{first-principles} calculations. Furthermore, we have also investigated the evolution of topological phases as a function of uniaxial strain. 
The presence of nodal line phases in unstrained Cu$_2$SnS$_3$ and the strain engineering that produces emergent phases, such as TFB, in this material prove its importance, as such features are rarely found. Studying the topological transport properties arising from the presence of TFB in this material will be an interesting avenue for future research.
Apart from this, the Weyl phase of Cu$_2$SnS$_3$ as a function of uniaxial strain in the presence of spin-orbit coupling (SOC) is also investigated. 
Our findings reveal the coexistence of seven Weyl points (three at $\Gamma$ and four at general positions) as well as nodal arcs, which have been lacking in earlier studies and are highly important for deepening our fundamental understanding of topological states of matter. 
In addition to this, it is also important to note that while type-I nodal line semimetals (NLSMs) have been reported in various materials, type-II NLSMs with only a single nodal ring are relatively rare. Moreover, 3D non-magnetic materials such as Mg$_3$Bi$_2$~\cite{10.1002/advs.201800897}, ZrSiSe~\cite{shao2020electronic}, and CrB$_4$~\cite{hou2024type} exhibit multiple nodal rings with type-II characteristics in the absence of SOC. 
But in the presence of SOC, Mg$_3$Bi$_2$ and ZrSiSe become trivial insulators, whereas CrB$_4$ transitions to a topological insulator. This suggests that the evolution of type-II NLSMs to the topological semimetallic phase not ubiquitous. The present material exhibits a transition from a type-II nodal-line semimetallic phase to Weyl phase in the presence of SOC.
Revealing such topological phases and their transitions in this material provides a new platform for investigating similar systems that host these features.
\begin{figure}[htbp]
    \centering
    \begin{subfigure}[b]{0.40\linewidth}
        \includegraphics[width=\linewidth, height=5.5cm]{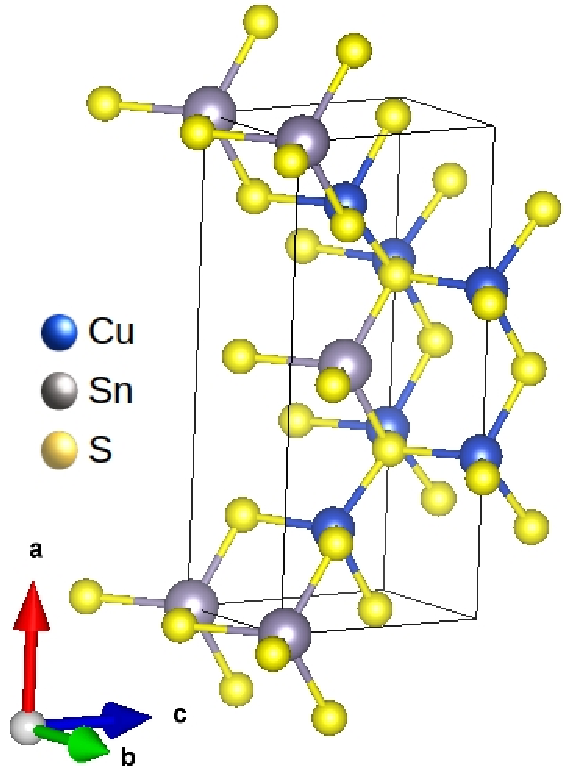}
        \caption{}
        \label{fig:lattice_structure}
    \end{subfigure}
    \hfill
    \begin{subfigure}[b]{0.55\linewidth}
        \includegraphics[width=\linewidth, height=5cm]{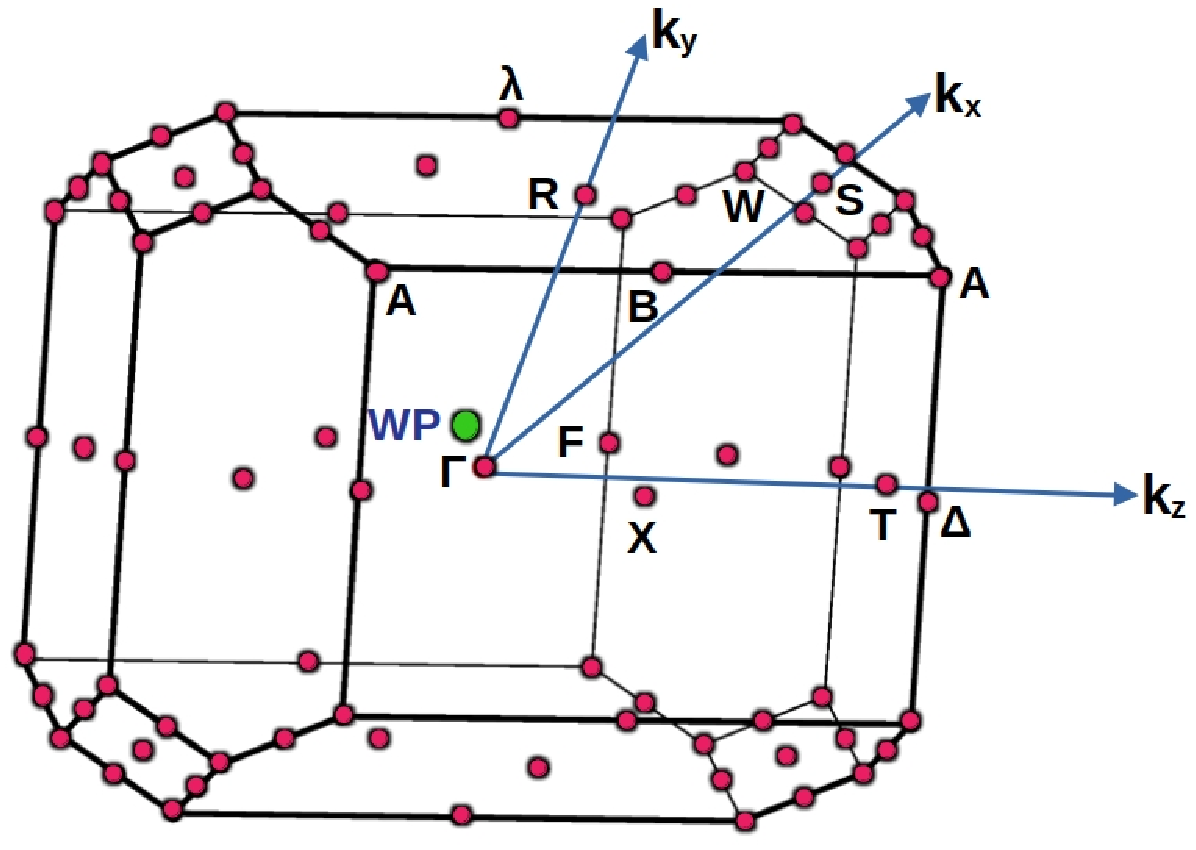}
        \caption{}
        \label{fig:BZ}
    \end{subfigure}
    \caption{\small (Color online) 
    (a) The crystal structure of Cu$_2$SnS$_3$. The blue, gray, and yellow balls denote Cu, Sn, and S atoms, respectively. Here a, b, and c correspond to the $x$-, $y$-, and $z$-axes, respectively. (b) The conventional Brillouin zone of Cu$_2$SnS$_3$.}
    \label{fig:lattice}
\end{figure}

\begin{figure*}\label{fig:band_a_axis_nosoc}
\includegraphics[width=0.24\linewidth, height=3.7cm]{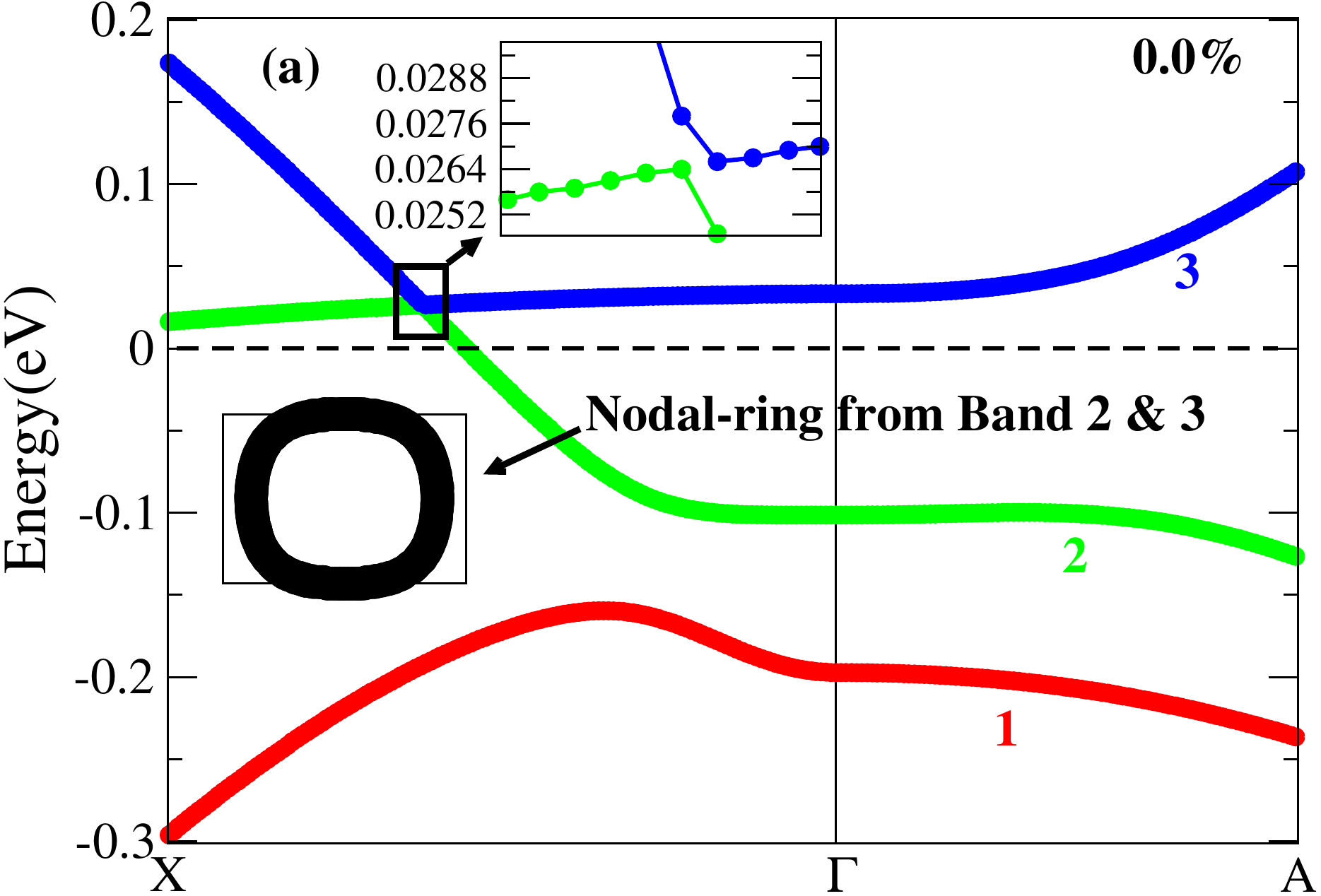}
\includegraphics[width=0.24\linewidth, height=3.7cm]{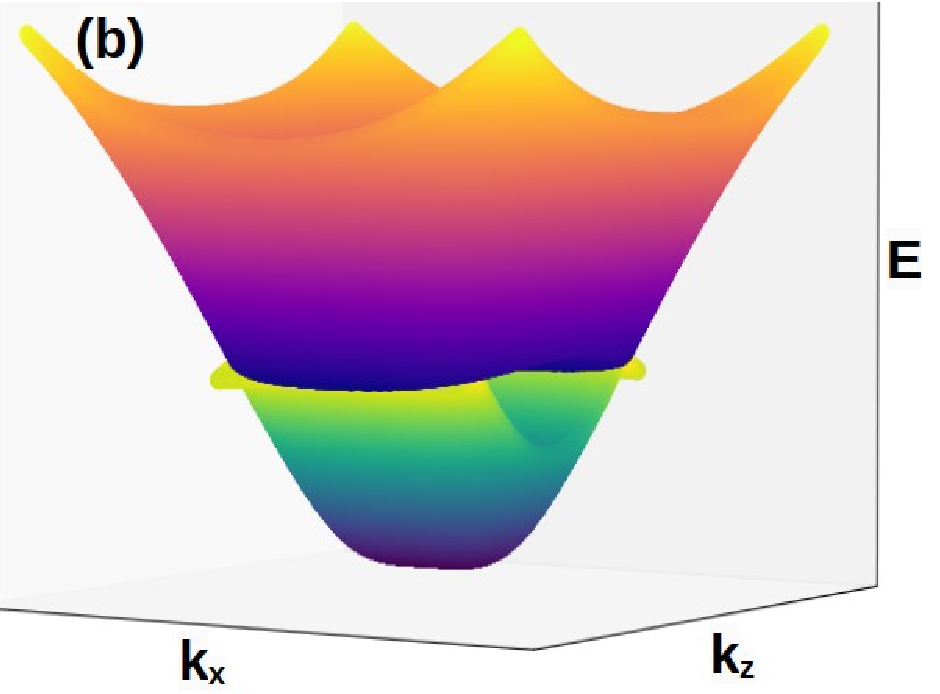}
\includegraphics[width=0.24\linewidth, height=4cm]{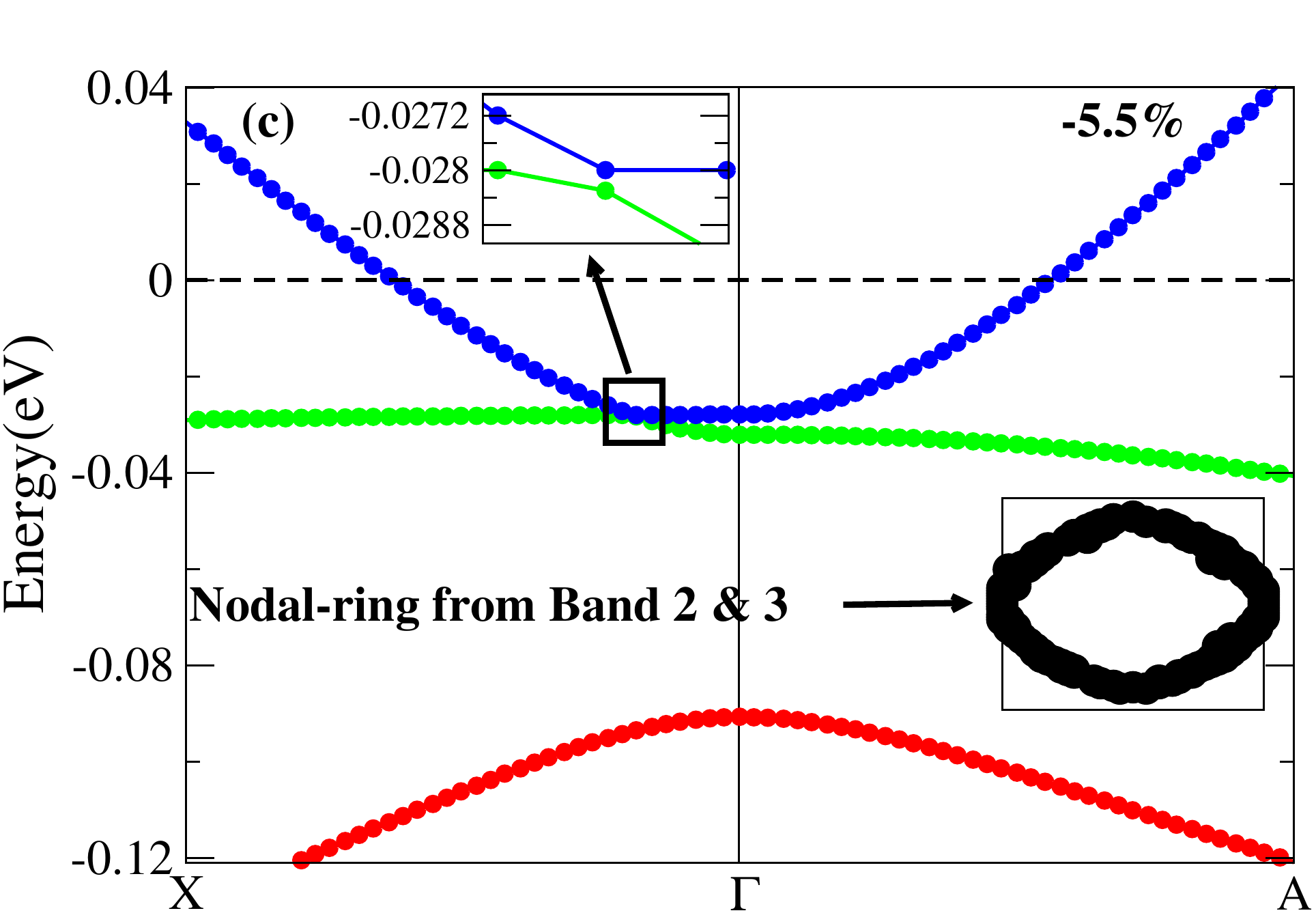}
\includegraphics[width=0.24\linewidth, height=4cm]{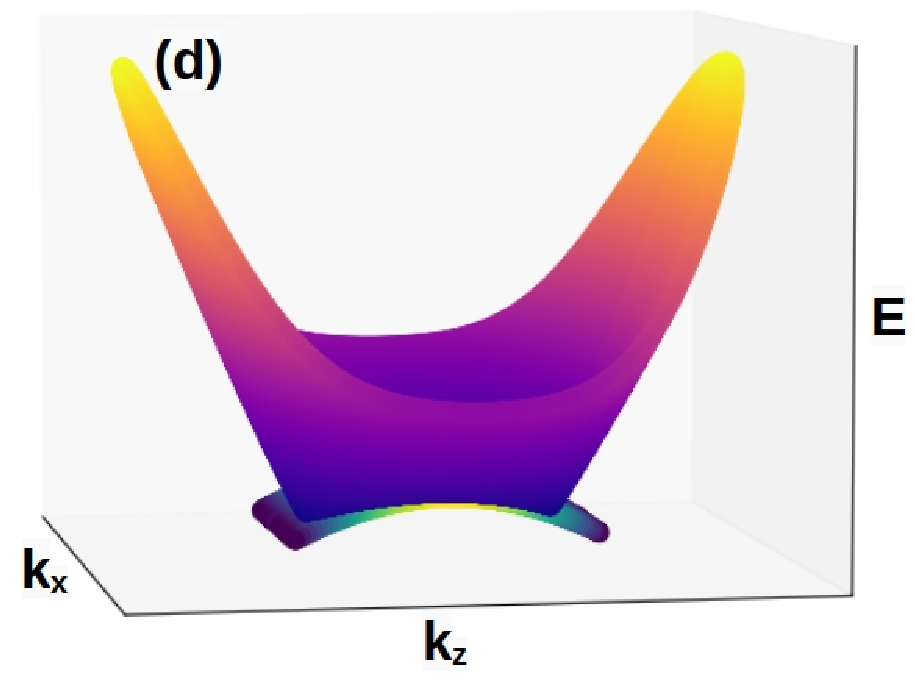}
\caption{\label{fig:band_a_axis_nosoc}\small{(Color online) 
Evolution of bulk band structure of Cu$_2$SnS$_3$ along high-symmetry points X-$\Gamma$-A in the first BZ under uniaxial strains in different magnitudes when SOC is excluded.
The negative sign in the figures stands uniaxial compressive strain (UCS) along the $a$ axis. The red, green and blue curves represent the band number 1, 2 and 3, respectively. Zero energy represents the Fermi level. (a) and (c) plots show the energy dispersions at 0\% and 5.5\% UCS, respectively. The subplot of (a) and (c) show the energy-gap bwtween band 2 and 3 along X-$\Gamma$ direction. We show the obtained topological nodal line in (a) and (c) at different magnitudes of strain in the first BZ apart from the X-$\Gamma$-A direction. (b)((d)) 3D representation of the energy (E) dispersions at 0 (-5.5)\% strain around type-II (type-III) nodal line in the $k_x$-$k_z$ plane. 
}}
\end{figure*}

\begin{figure*}\label{fig:band_a_axis_soc}
\includegraphics[width=0.40\linewidth, height=5.8cm]{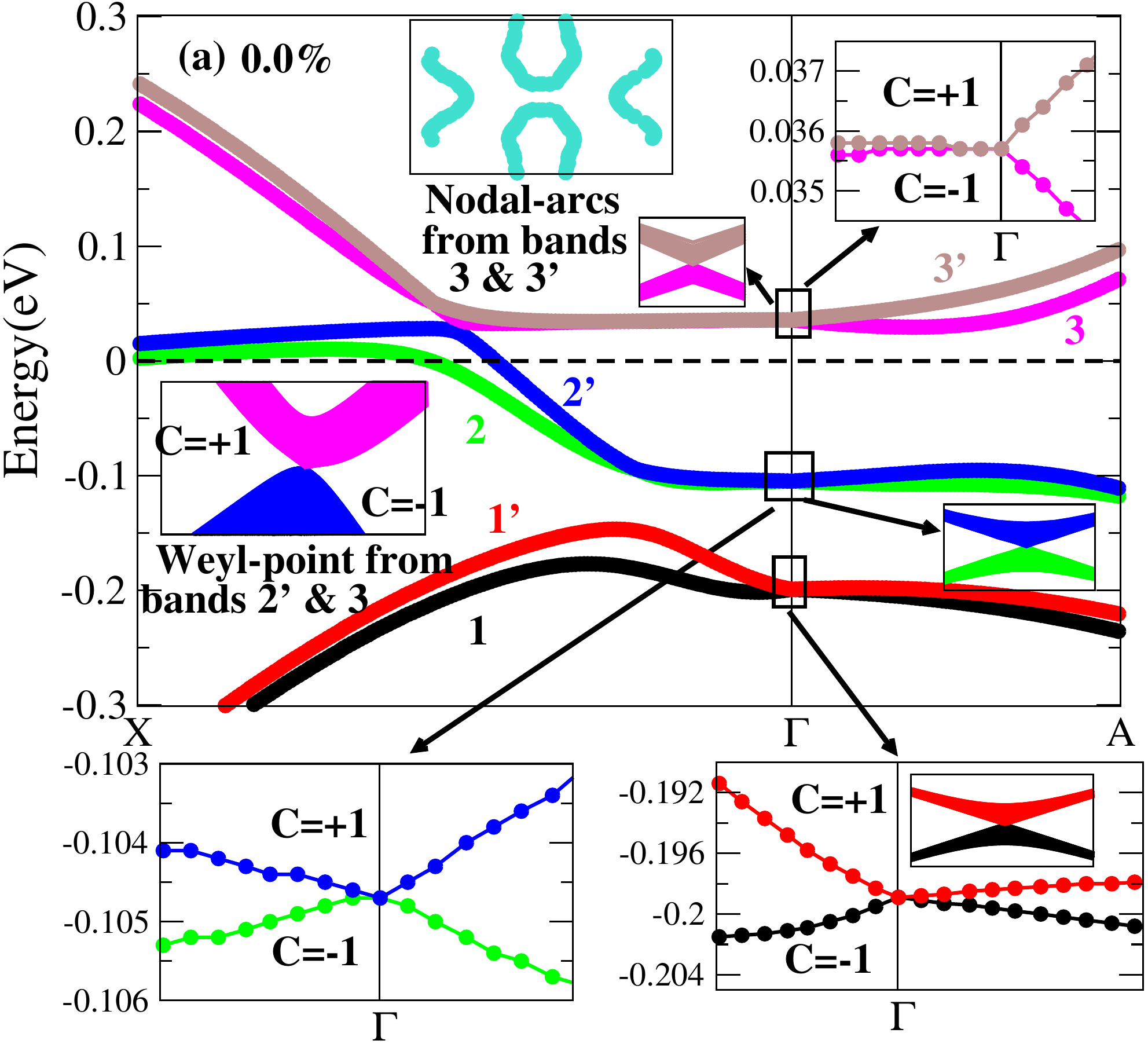} 
\includegraphics[width=0.40\linewidth, height=6.0cm]{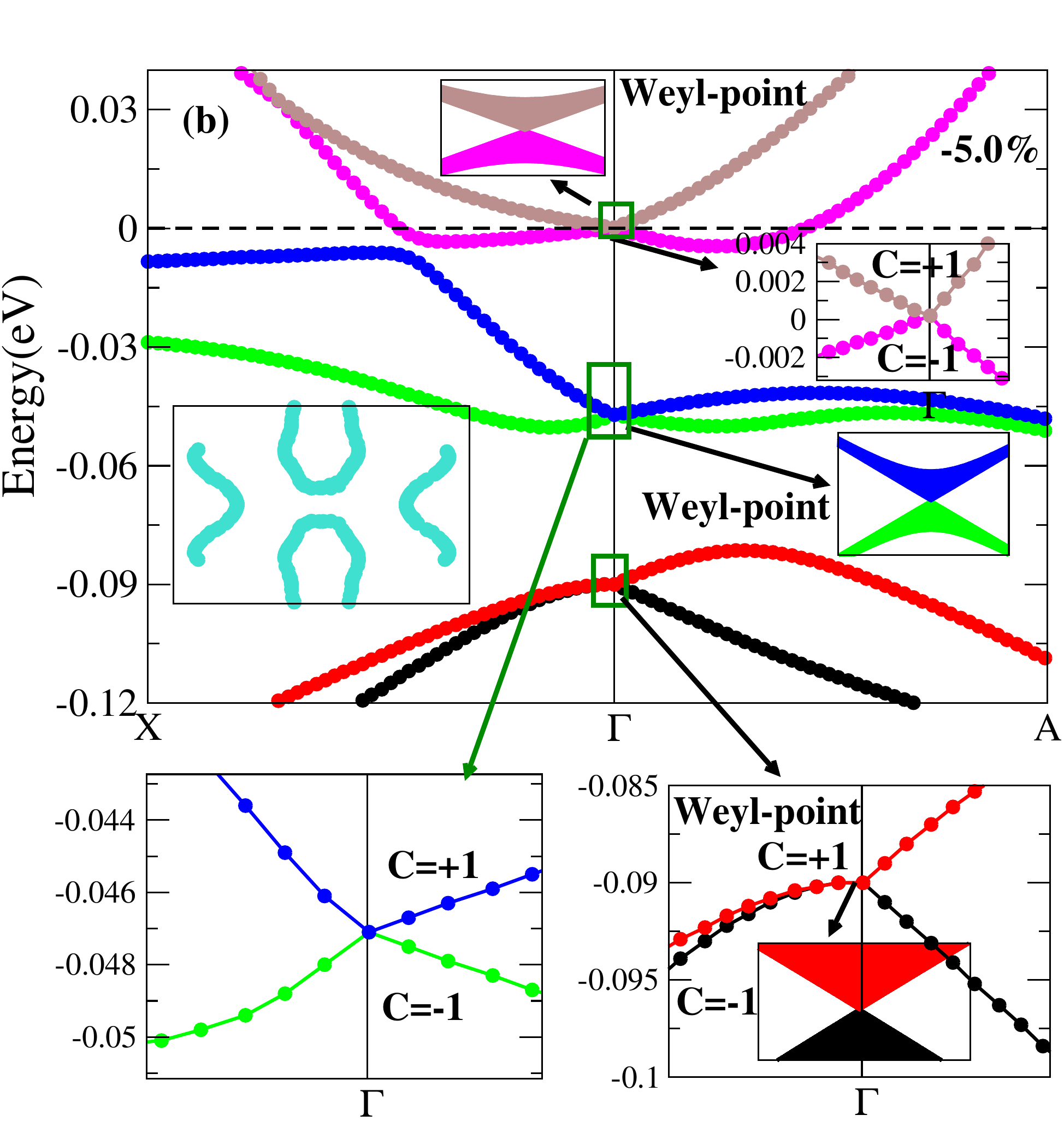}
\caption{\label{fig:band_a_axis_soc}\small{(Color online) 
Evolution of bulk band structure of Cu$_2$SnS$_3$ along high-symmetry direction X-$\Gamma$-A in the first BZ under uniaxial strains in different magnitudes when SOC is included. The negative sign in the figures stands UCS along the $a$ axis. The black, red, green blue, magenta and brown curves represent the band number 1, $1^\prime$, 2, $2^\prime$, 3 and $3^\prime$, respectively. (a) and (b) plots show the energy dispersions at 0\% and 5.5\% UCS. We show the zoom-up of the touching of different combinations of bands. The chiralities (C) for each band are also indicated in the plots. The coordinates of the Weyl point between bands $2^\prime$ and 3 are ($\pm\frac{2\pi}{a}$(0.226), $\pm\frac{2\pi}{b}$(0.004), 0.000), and the energy dispersion around the obtained Weyl point is shown in subplot (a). We also show the nodal arc in the subplot of Fig. (a) and (b).
}}
\end{figure*}

\begin{figure*}\label{fig:surface_state}
\includegraphics[width=0.32\linewidth, height=4.0cm]{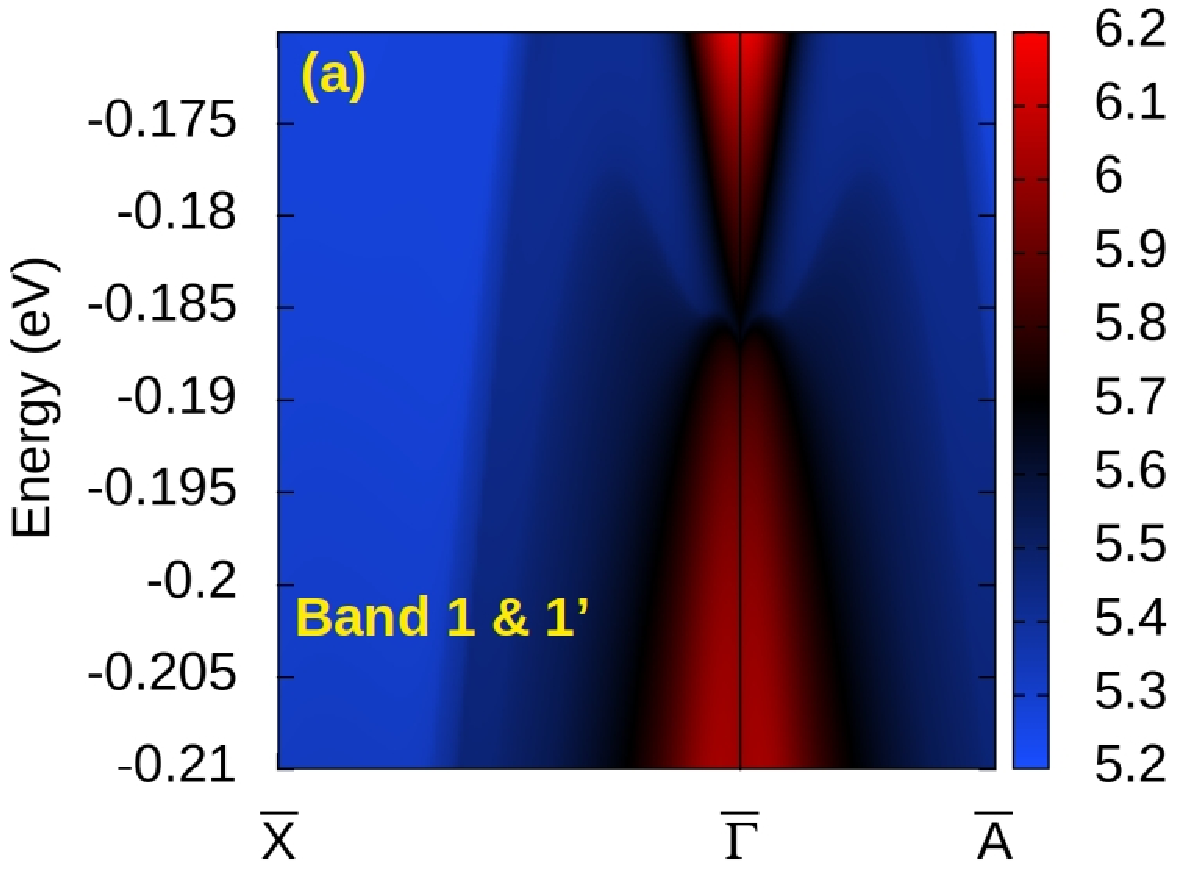}
\includegraphics[width=0.32\linewidth, height=4.1cm]{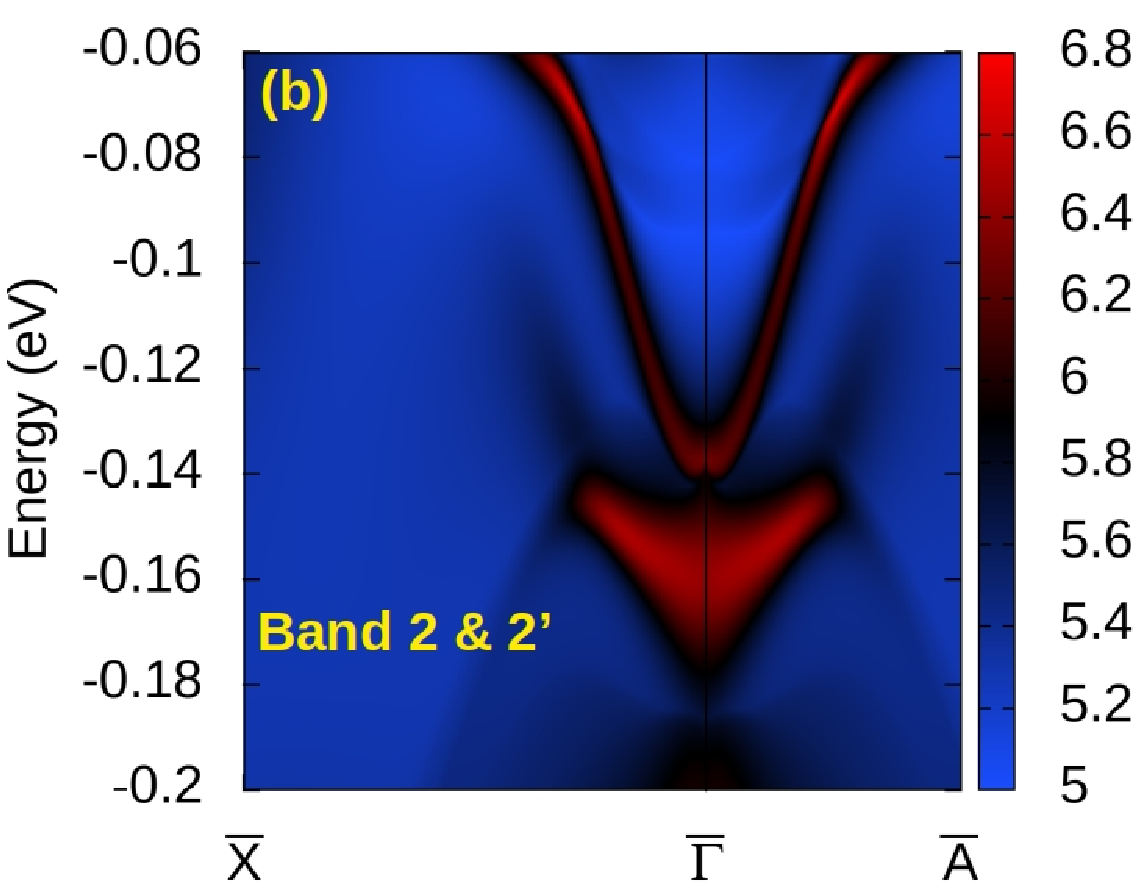}
\includegraphics[width=0.32\linewidth, height=4.0cm]{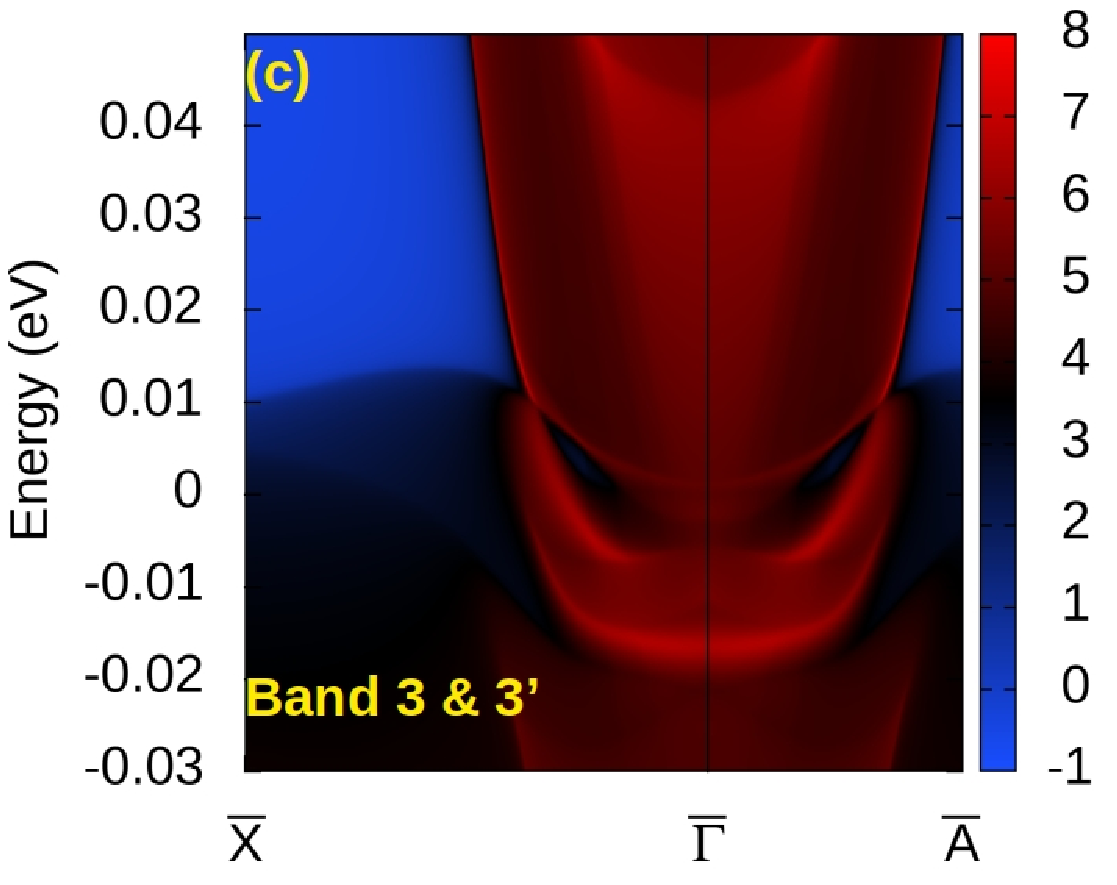}
\includegraphics[width=0.32\linewidth, height=4.5cm]{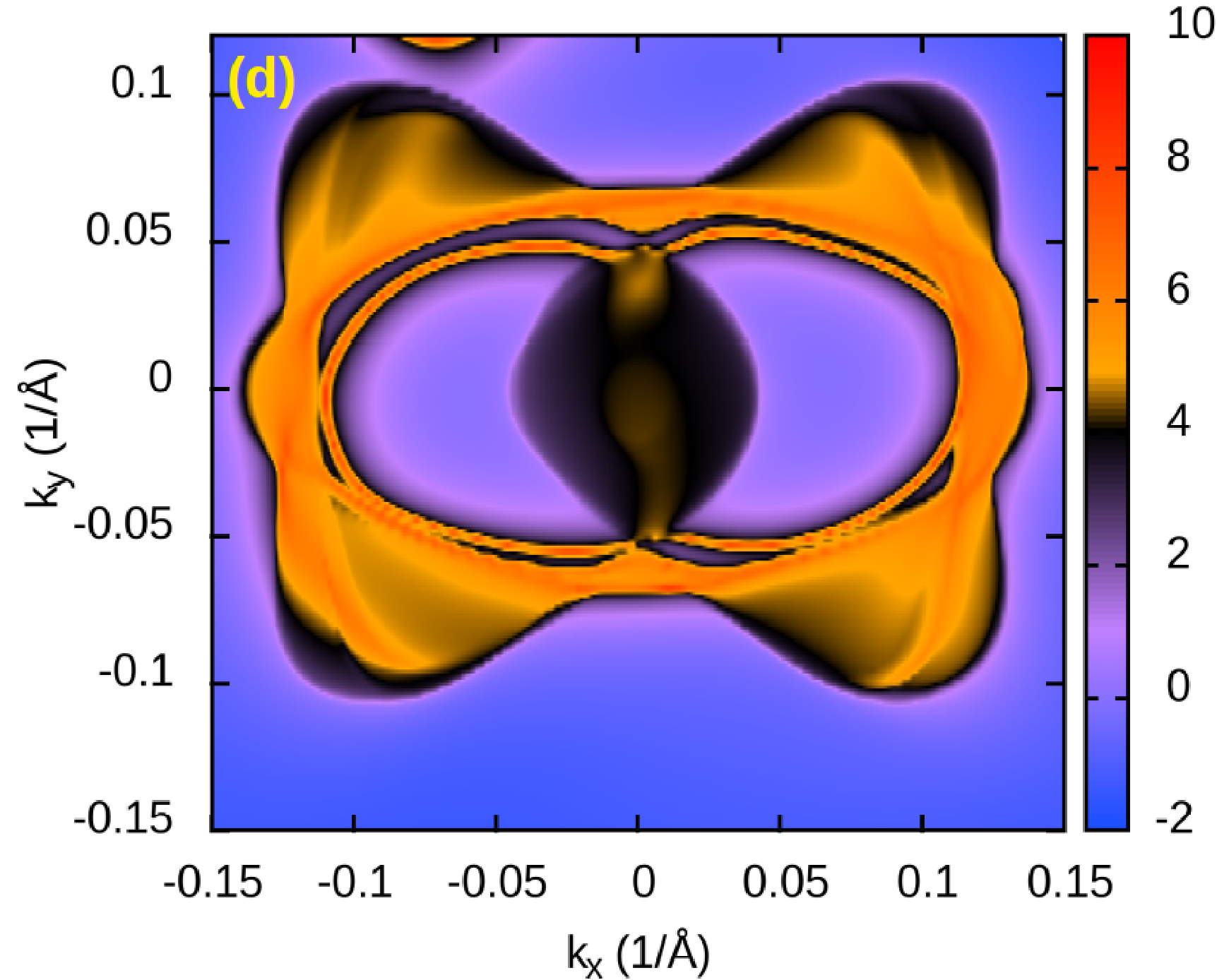}
\includegraphics[width=0.32\linewidth, height=4.5cm]{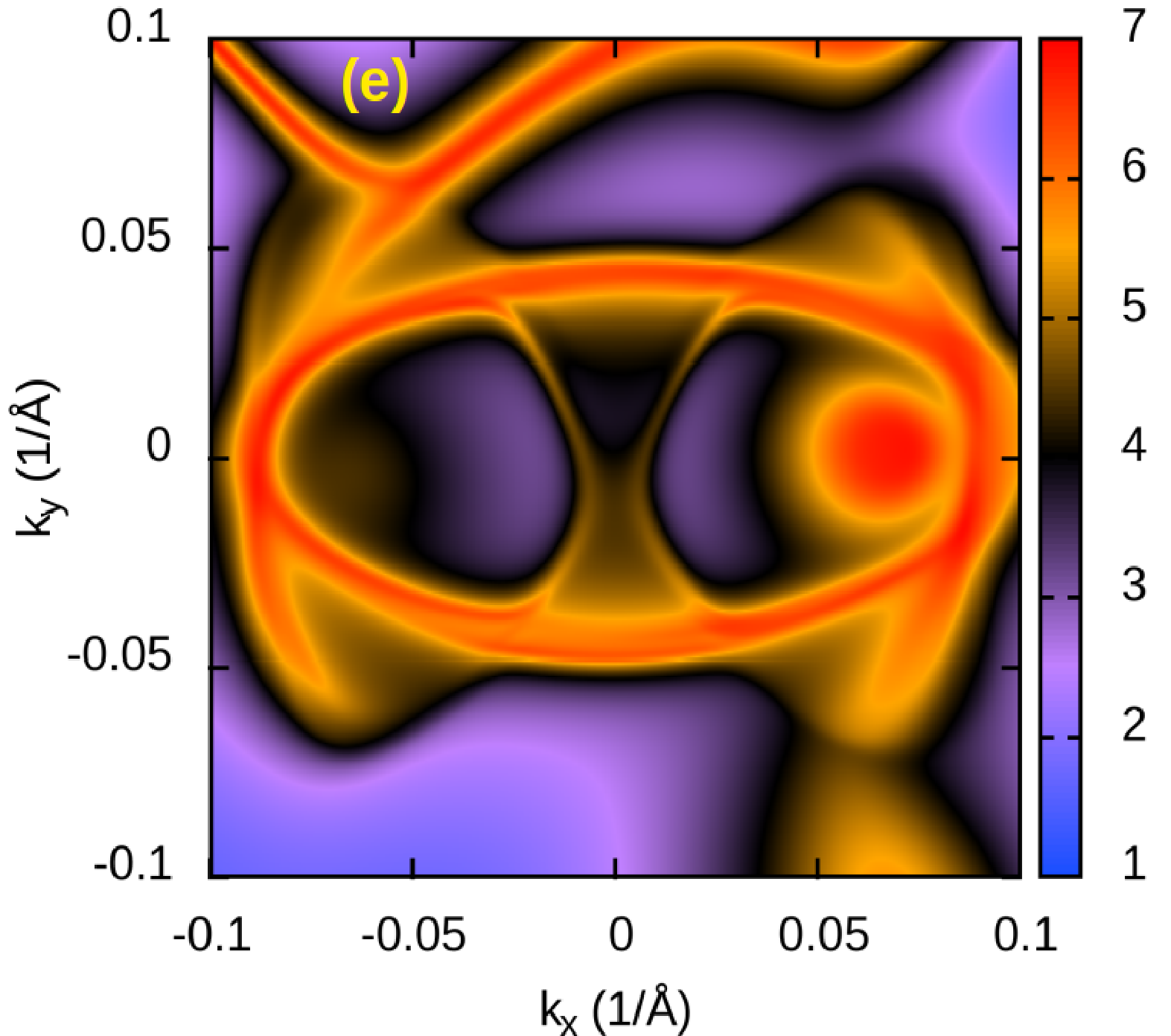}
\includegraphics[width=0.32\linewidth, height=4.5cm]{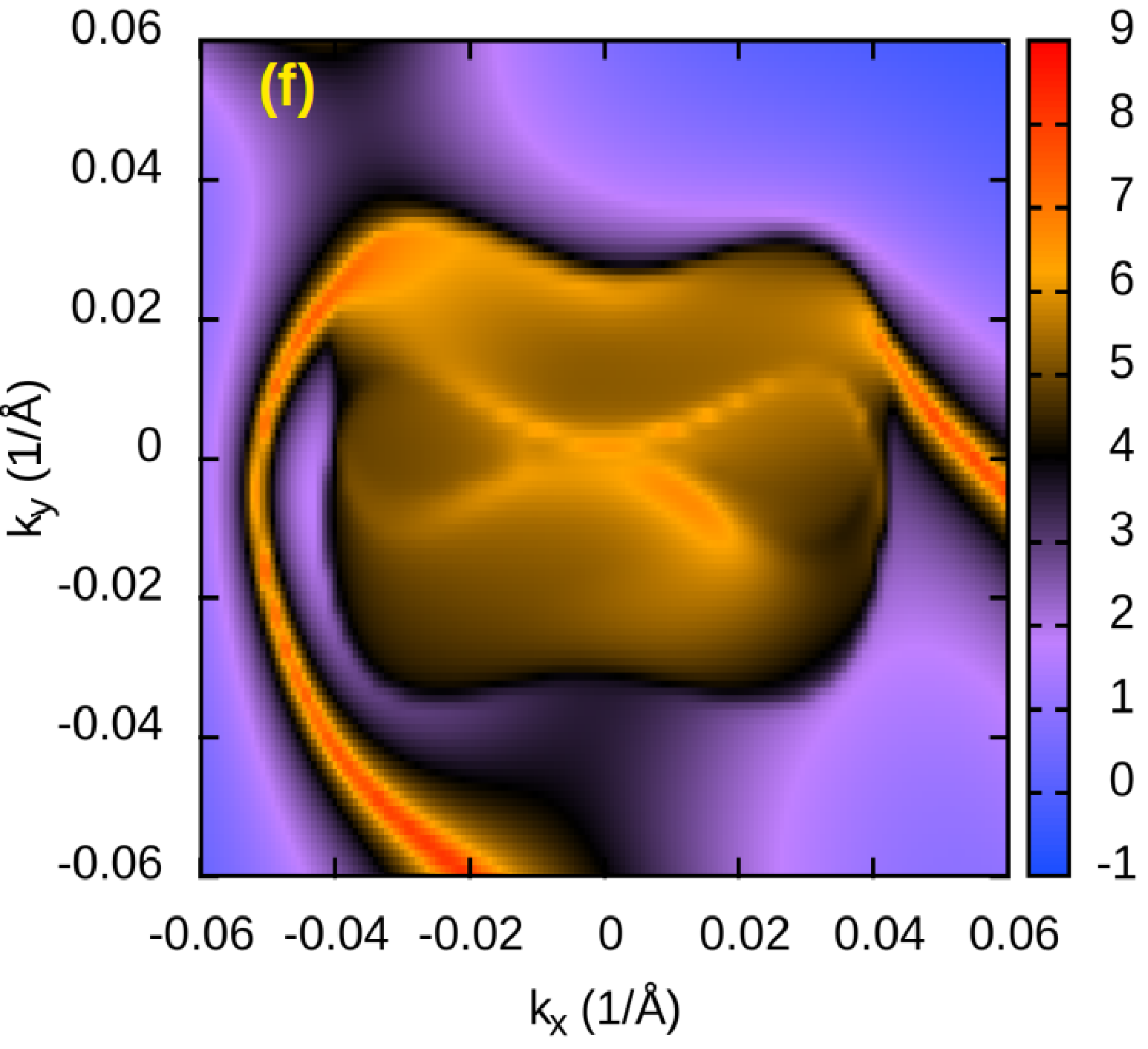}
\caption{\label{fig:surface_state}\small{(Color online) 
Surface density of states for the (001) surface (a) band 1-$1^\prime$, (b) band 2-$2^\prime$, and (c) band 3-$3^\prime$ of Cu$_2$SnS$_3$ at 0\% strain along two dimensional high-symmetric directions. 
(d)-(f) plot show non-trivial topological Fermi arcs of Cu$_2$SnS$_3$ of (001) surface for band 1-$1^\prime$ at energy -185 meV, band 2-$2^\prime$ at energy -104.5 meV and band 3-$3^\prime$ at energy 35.5 meV for the top surface of the unit cell.}}
\end{figure*}


{\it Computational details.|}
Using the \textit{first-principles} approach, we have used the \texttt{WIEN2k} package~\cite{blaha2020wien2k} to calculate the ground state energy of orthorhombic Cu$_2$SnS$_3$. 
The generalized gradient approximation (GGA)~\cite{PhysRevLett.77.3865} within the Perdew-Burke-Ernzerhof is preferred as an exchange-correlation (XC) functional. This XC functional is used to optimize the geometry and lattice constant. The optimised lattice parameters are $a$=11.61 \AA, $b$=3.92 \AA, and $c$=5.43 \AA. The relaxed wyckoff positions of Cu, Sn, S1, and S2 are (0.8297, 0.5, 0.49053), (0.0, 0.0, 0.018978), (0.82306, 0.0, 0.26436), and (0.5, 0.0, 0.22018), respectively. A 12$\times$12$\times$12 $k$-mesh size is used for the self-consistent calculations. Furthermore, the energy convergence limit for the self-consistent method is set to $10^{-7}$ Ry/cell for the calculations. 
The analysis of Weyl nodes and nodal lines was performed using our in-house code, \texttt{PY-Nodes}~\cite{PANDEY2023108570}, which is based on the Nelder-Mead function minimization method~\cite{10.1093/comjnl/7.4.308, PANDEY2024109281}. 
To calculate the topological charge or chirality for each Weyl point, we have used the \texttt{WloopPHI} code~\cite{saini2022wloopphi}. 
For calculating the chirality, we have taken 250 Wilson loops with a 12$\times$12$\times$12 k-mesh size. We construct the Wilson loop around the Weyl point by choosing four vertex points to define the loop, and then find the evolution of the Berry curvature along this trajectory.
To investigate topological properties such as surface states and Fermi arcs, a tight-binding (TB) model based on maximally localized Wannier functions (MLWF) is obtained using the \texttt{WANNIER90} code~\cite{mostofi2008wannier90}. The surface state calculations are performed using the \texttt{WANNIERTOOLS} package~\cite{WU2017}.

{\it Material realization.|}
Recently, the symmorphic $Imm2$-phase of the Cu$_2$SnS$_3$ class of materials has been theoretically reported as a new type of topological semimetal. The lattice structure and the corresponding Brillouin zone of the $Imm2$ phase of Cu$_2$SnS$_3$ are shown in Figs. \ref{fig:lattice} (a) and \ref{fig:lattice} (b), respectively.
In particular, Cu$_2$SnS$_3$ exhibits a diverse range of phases, including monoclinic, cubic, tetragonal, hexagonal, and orthorhombic, associated with the $Cc$, $F\bar{4}3m$, $I\bar{4}2m$, $P6_3/mmc$, and $Imm2$ space group symmetries, respectively~\cite{ONODA20001563, CHEN1998144, 10.1063/1.2790491, pssc.200982746, PhysRevResearch.4.033067}. The $Cc$ and $F\bar{4}3m$ phases of this compound have been synthesized and systematically studied for their electronic and vibrational properties through experiments and \textit{first-principles} calculations. 
The electronic and phononic topological transport properties of the $Imm2$ phase of this material remain unexplored from both experimental and theoretical perspectives. Therefore, it is necessary to first study their electronic topological properties in the absence and presence of strain.

{\it In the absence of SOC.|} The electronic band structure of Cu$_2$SnS$_3$ in Fig. \ref{fig:band_a_axis_nosoc} (a) displays important features. First, we find that the band touching (2-3) appears along X-$\Gamma$ high-symmetry path with a linear dispersion at the energy $\sim 26$ meV, which is above the Fermi level. 
{However, a zoomed-in scale of the figure shows a gap of $\sim$1.3 meV between bands 2 and 3. This indicates that along X-$\Gamma$ high-symmetry path, topological quantum phases are absent. Therefore, it is necessary to search for band-touching points in the entire Brillouin zone (BZ). Consequently, we have searched for band-touching points in the entire BZ and found that they form a nodal line.} 
The obtained nodal line lies in the $k_x$-$k_z$ plane with $k_y$=0, as shown in the subplot of Fig. \ref{fig:band_a_axis_nosoc} (a). 
The nodal line formed by band-touching points lies at ($k_x$,0,$k_z$). Since the studied material belongs to the space group $Imm2$, it possesses mirror reflection symmetries $M_x$ and $M_y$, as well as a twofold rotation symmetry $C_{2z}$, which is part of the little group. The $k_y$=0 plane is invariant under the mirror symmetry $M_y$, i.e., ($k_x$, $k_y$, $k_z$) $\rightarrow$ ($k_x$, -$k_y$, $k_z$). The $M_x$ \& $C_{2z}$ symmetries transform a point as ($k_x$, $k_y$, $k_z$) $\rightarrow$ (-$k_x$, $k_y$, $k_z$) and ($k_x$, $k_y$, $k_z$) $\rightarrow$ (-$k_x$, -$k_y$, $k_z$), respectively. These operations allow for band crossings.
To determine the nature of the obtained nodal line, we calculated the energy spectrum around the obtained node points. The corresponding results are shown in Fig. \ref{fig:band_a_axis_nosoc} (b). The 3D representation of the energy spectrum with respect to the wave vector \textbf{k} reveals that both bands have identical slopes. Such band-dispersion is commonly referred to as type-II~\cite{He_2018}. This indicates that the low-energy excitations in Cu$_2$SnS$_3$ may correspond to type-II nodal line fermions in the absence of SOC~\cite{soluyanov2015type, 10.1126/sciadv.1603266}. To verify whether the touching points in the 3D representation of the energy spectrum indeed form a nodal ring, we calculated the 2D projection where the two bands are degenerate at the \textbf{k}-points (see Fig. 1 and corresponding text in the SM~\cite{supplementry}).
The figure clearly indicates that the bands form a nodal ring in the $k_x$-$k_z$ plane with $k_y = 0$. Therefore, we conclude that Cu$_2$SnS$_3$ is a type-II nodal line semimetal.

Strain has proven to be a powerful and effective approach for engineering the electronic properties of quantum materials by tuning their band structures. Here, we investigate the impacts of strain on the dispersion curves, which directly influence the topological properties of quantum materials. Considering the directionally dependent properties, we applied uniaxial compressive strains (UCS) along the \textit{a} axis ($x$-axis), ranging from 0\% to 6\%, which are achievable in experiments.  
We have calculated the electronic dispersion for each strained structure of Cu$_2$SnS$_3$ in the absence of SOC and observed how the dispersion evolves with different magnitudes of strain.
It is found that at 5\% compressive strain, the band-touching points come very close to the Fermi level (see Fig. 2 and corresponding text in the SM~\cite{supplementry}).
At this strain (5\%), the bands behave such that in some directions within the $k_x$-$k_z$ plane, both bands share the same slope, while in other directions, one band has zero slope and the other has a non-zero slope. We refer to this situation as quasi-type III, where both of these features coexist.
It is also important to mention that at this strain, the quasi-type III feature appears to be accidental resulting from both bands come close the Fermi energy. Since at this strain the lattice parameters are $a\neq b\neq c$ as in the unstrained structure, and the system still belongs to an orthorhombic body-centered lattice, this means that the quasi-type feature is therefore not due to appearance of extra symmetry at this strain.
In addition to this, the area covered by the nodal line decreases as the compressive strain value increases (see details in the SM~\cite{supplementry}).  
With the further application of a 5.5\% compressive strain, the situation becomes markedly different from the unstrained structure. At this point, quasi-type-III dispersion transitions into type-III, which results in the formation of type-III nodal line. The corresponding band structure and 3D representation of the energy spectrum are shown in Figs. \ref{fig:band_a_axis_nosoc} (c)-(d). From Fig. \ref{fig:band_a_axis_nosoc} (d), it is seen that one of the energy bands is dispersionless across the entire $k_y$=0 plane. This type of energy band-dispersion is generally referred to as type-III. Moreover, we have also plotted the projection of the two touching 3D dispersion bands onto $k_x$-$k_z$ plane, where the intersection forms a nodal line (see Fig. 3 and corresponding text in the SM~\cite{supplementry}). From the above discussion, it is revealed that at 5.5\% strain, Cu$_2$SnS$_3$ exhibits a type-III nodal line. Furthermore, with the application of a 5.6\% compressive strain, the type-III nodal line transforms into Weyl points (see Fig. 4 and corresponding text in the SM~\cite{supplementry}). 
This is another interesting result that the band topology changes after 5.5\% compressive strain, leading to the appearance of Weyl nodes. To confirm whether it is indeed a Weyl node at this strain value, we have calculated the chirality and found it to be $\pm$1. 
Thus, the obtained results reveal the presence of several topological phases under compressive strain ranging from 0\% to 5.6\%, including a type-II nodal line at 0\%, a quasi-type-III nodal line phase at 5\%, a type-III nodal line phase at 5.5\%, and a Weyl point at 5.6\%. 
To determine the nature of this Weyl point, we calculated the 3D energy spectrum around the node point and found that one energy band is dispersionless, while the other is dispersive across the entire $k_x$=0 plane. This type of energy band dispersion is generally referred to as a type-III feature of the Weyl point (see Fig. 4 in the SM~\cite{supplementry}).

Along with this, we have also calculated the density of states (DOS) for different values of uniaxial compressive strain (0\%, 5\%, 5.5\%, and 5.6\%). The corresponding plot is shown in Fig. 5 of the SM~\cite{supplementry}. Firstly, we find that the DOS within the studied range of UCS (0\%$\leq$UCS$\leq$ 5.6\%) remains non-vanishing at the Fermi level. Secondly, at 0\% strain, it is evident from the figure that the DOS exhibits almost parabolic in nature around the Fermi level. With increasing strain magnitude up to 5.6\%, this parabolic behavior gradually transitions into a nearly flat DOS above the Fermi level, forming a plateau that extends over an energy range of $\sim$ 60 meV. Since the DOS includes contributions from the full Brillouin zone, whereas the flat band is seen only along the X-$\Gamma$-A directions and hence DOS may not fully capture the flat-band nature. However, our 3D band dispersion at 5.6\% UCS clearly reveals flat-band characteristics along the X-$\Gamma$-A directions near the $\Gamma$ point. Moreover, it is well known that the group velocity $v_g=1/\hbar (\partial k/\partial E)$, the flat-band slope, where electrons barely change energy with momentum corresponds to a very low group velocity, implying low kinetic energy, reduced electronic dispersion, and consequently, enhanced electron-electron interactions. It is typically seen that electron-electron interactions may lead to correlation-driven phenomena such as enhanced effective mass, charge ordering, or Mott-like behavior~\cite{hu2023correlated, chen2024emergent, karmakar2025}.
Therefore, the application of external strain enables the realization of multiple topological phases, offering a promising platform for exploring non-trivial physical phenomena. In experiments, these topological phase transitions can be achieved by applying appropriate uniaxial compressive stress. 

It is also important to highlight here that the nodal line has been found to be useful for various applications, and there are different ways to reduce the strength of SOC in real materials. For example, SOC strength can be tuned by chemical substitution, the application of external electric and magnetic fields~\cite{schott2017tuning, 10.1063/1.4804334, fan2016electric}. 
Keeping this in mind we have studied the robustness of nodal line in this material by varying the strength of SOC. For this purpose, we performed electronic structure calculations with different SOC strengths. Since the SOC Hamiltonian is inversely with the square of the speed of light, decreasing the speed of light ($c$) effectively enhances the SOC strength. Accordingly, $c$ is treated as a tunable parameter and reduced to increase the SOC strength in our calculations. We reduced $c$ from 1370.3 to 548.1 a.u., which corresponds to an increase in SOC strength from $\sim$0.0 to 2.0. It is important to note that the SOC strength is evaluated from the difference in the total energy of the Cu$_2$SnS$_3$ system calculated with and without the inclusion of SOC. We find that at $\sim$0.0 meV SOC strength, Cu$_2$SnS$_3$ exhibits a nodal-line phase. The nodal-line phase persists as the SOC strength increases from 0.0 up to $\sim$2.0 meV. Interestingly, above 2.0 meV, the nodal-line phase evolves into a Weyl phase. This indicates that 2.0 meV represents a critical SOC threshold, beyond which the nodal-line phase vanishes and the Weyl phase emerges. The Weyl phase remains stable up to an SOC strength of 82.56 meV, corresponding to $c$=137.03 a.u.

Since, in Cu$_2$SnS$_3$, the lattice parameter along \textit{a} axis is larger than the other two. As a result, growing single crystals along \textit{a} direction (referred to as the easy axis) is relatively simple and often preferred due to its alignment with the longest lattice parameter and the resulting lower-energy configurations. Therefore, uniaxial compressive strain of up to 6\% along the \textit{a} axis may be experimentally achievable. In order to confirm that the orthorhombic structure remains intact under up to 6\% UCS along the \textit{a}-axis, we calculated the elastic constants and verified them against the elastic stability conditions for an orthorhombic system ~\cite{PhysRevB.90.224104}. The calculated elastic constants at 6\% UCS are as follows (in GPa):

\[
\begin{aligned}
C_{11} &= 167.58, & C_{22} &= 109.96, & C_{33} &= 96.14,\\
C_{44} &= 3.52,   & C_{55} &= 36.33,  & C_{66} &= 60.04,\\
C_{12} &= 16.84,  & C_{13} &= 65.16,  & C_{23} &= 50.05.
\end{aligned}
\]
Our calculated values satisfy the elastic stability conditions for an orthorhombic system, which are:  
     (i) $C_{ii} > 0$, (ii) $C_{11}C_{22} > C_{12}^2$,
     (iii) $C_{11}C_{22}C_{33} + 2C_{12}C_{13}C_{23} - C_{11}C_{23}^2 - C_{22}C_{13}^2 - C_{33}C_{12}^2 > 0$,
     (iv) $C_{ii} + C_{jj} - 2C_{ij} > 0$, and
     (v) $C_{11} + C_{22} + C_{33} + 2(C_{12} + C_{13} + C_{23}) > 0$.

The above discussion reveals that electronic dispersions are highly sensitive to mechanical strain. At this point, we focus on another interesting feature: the topological three-dimensional flat band, which is highly non-trivial because it is typically reported in special types of lattices such as kagome and Lieb or twisted bilayer/trilayer graphene~\cite{PhysRevB.110.024504, PhysRevB.109.L041109, PhysRevB.99.235118, MA202118, PhysRevB.110.195112}. However, when applying a 5.6\% UCS on Cu$_2$SnS$_3$, we reveal a topological flat band in the system, which is reported in an orthorhombic body-centered lattice. Moreover, hydrostatic pressure and uniaxial strain are powerful methods to achieve topological flat bands. For example, earlier studies~\cite{PhysRevResearch.4.023114} on superlattices of the HgTe class of materials (with space group F4$\bar{3}$m) demonstrated how topological phases evolve as a function of hydrostatic pressure and uniaxial strain. Therefore, the emergence of a topological flat band through uniaxial strain is highly significant, as this approach allows achieving the feature simply by applying strain, without the need to create superlattices or introduce disorder into the system.

In addition to this, we have also investigated the role of atomic orbitals, as they are responsible for driving various topological phases. The orbitals contributing to the topmost valence band (TVB - Band 2) and the bottommost conduction band (BCB - Band 3) along the $k_x$-$k_z$ plane ($k_y$=0), where the nodal line transitions from type-II to type-III under the application of UCS, have been analyzed. It is found that the TVB and BCB are primarily contributed by Cu-\textit{3d} and S-\textit{3p} orbitals. For the unstrained structure, it is found that across the entire $k_x$-$k_z$ plane, the Cu-\textit{3d} orbitals contribute more to TVB than the S-\textit{3p} orbitals. However, in the BCB, the Cu-\textit{3d} and S-\textit{3p} orbitals exhibit complementary periodic contributions  where the Cu-\textit{3d} orbital contribution is higher, the S-\textit{3p} contribution is lower, and vice versa. For a clearer understanding, we have also calculated the character contribution to the TVB and BCB for the \textbf{k}-points inside the nodal loop. It is found that the \textbf{k}-points lying inside the nodal ring have a more character contribution from Cu-\textit{3d} orbitals to TVB than from S-\textit{3p} orbitals, whereas for BCB, the S-\textit{3p} orbitals contribute more than the Cu-\textit{3d} orbitals. Similarly, \textbf{k}-points lying outside the nodal ring have a more character contribution from Cu-\textit{3d} orbitals to TVB than from S-\textit{3p} orbitals, whereas for BCB, the S-\textit{3p} orbitals contribute more than the Cu-\textit{3d} orbitals. This inversion in the band character is commonly referred to as an intraband \textit{d-p} inversion. Such band inversion is reported to facilitate the formation of nodal line phase~\cite{Pandey_2023}. Similarly, we have also studied the origin of the strain-induced type-III nodal line \& Weyl phase and found similar results.


{\it In the presence of SOC.|}
Spin-orbit coupling (SOC) interactions are a key ingredient for generating non-trivial topological properties. In Cu$_2$SnS$_3$, the effect of SOC plays a vital role in exhibiting topologically non-trivial properties. 
SOC splits band 2 (3) into 2 and $2^\prime$ (3 and $3^\prime$) along the X-$\Gamma$-A high-symmetry path, as shown in Fig. \ref{fig:band_a_axis_soc} (a). 
However, the degeneracy is preserved between bands $2^\prime$ and 3. This accidental degeneracy leads to a Weyl point with chirality (C) $\pm$1. The co-ordinates of the obtained Weyl point are ($\pm\frac{2\pi}{a}$(0.226), $\pm\frac{2\pi}{b}$(0.004), 0.000). 
Furthermore, band 1 splits into 1 and $1^\prime$, resulting in a Weyl point near -198 meV at the $\Gamma$ point. Similarly, band 2 (3) splits into bands 2-$2^\prime$ (3-$3^\prime$), which also produces Weyl points near -104.5 (35.5) meV, at the $\Gamma$ point. Notably, another interesting feature is observed: the presence of a nodal-arcs between bands 3 and $3^\prime$, as shown in the inset of Fig. \ref{fig:band_a_axis_soc} (a). Therefore, in the unstrained structure with SOC included, the coexistence of both Weyl nodes and nodal lines distinguishes this material from others. Interestingly, such a coexistence has not been reported in previous studies~\cite{PhysRevResearch.4.033067}. 
The possible reason for this may be the use of MLWFs, which are employed to construct the TB model for predicting the topological quantum phases of the material. To obtain an accurate TB Hamiltonian, good Wannier fitting is essential because the bands and projectors play a significant role in achieving a reliable TB Hamiltonian. It is also important to keep in mind that highly accurate results may not be expected from an empirical tight-binding model~\cite{vanderbilt2018berry}. Therefore, there is a high chance of missing important topological quantum phases present in the material.
Since the Weyl and nodal-arc phases are found between bands 3-$3^\prime$, which lie above the Fermi level. To make these features experimentally accessible, experiments should be specifically designed to probe the states above the Fermi level. Indeed, various studies suggest that such features, including Weyl points or nodal lines, can be observed by tuning the chemical potential via electron doping. In the following references such attempts have been made~\cite{cheng2024tunable, science.aav2873}.
From Fig. \ref{fig:band_a_axis_soc} (b), an interesting and important result is observed: upon applying 5\% UCS, the degeneracy of bands $2^\prime$ and 3 is lifted, creating a gap of 5.86 meV. However, for UCS up to $<$5\%, the Weyl phase remains robust, but the coordinates of the Weyl point shift from ($\pm\frac{2\pi}{a}$(0.226), $\pm\frac{2\pi}{b}$(0.004), 0.000) to ($\pm\frac{2\pi}{a}$(0.099), $\pm\frac{2\pi}{b}$(0.003), 0.000). This indicates that the presence of Weyl points in the unstrained structure is due to accidental degeneracy between bands $2^\prime$ and 3, which is lifted upon applying UCS $\geq$5\%. 
Similar studies are done with varying magnitudes of UCS, and we found similar results for each strain value (see Fig. 6 and corresponding text in the SM~\cite{supplementry}). Also, for each values of strain up to $<$5\%, we found only four Weyl points, indicating the minimal number of Weyl points that a system with time-reversal invariance can host (see details of coordinates of the Weyl points as mentioned in Table 1 of the SM~\cite{supplementry}).
It is important to note that, among the seven Weyl points, four Weyl points [($\pm\frac{2\pi}{a}$(0.226), $\pm\frac{2\pi}{b}$(0.004), 0.000)] are formed between bands $2^\prime$ and 3 due to accidental degeneracy, which becomes gapped upon applying 5\% UCS. In contrast, the Weyl point at $\Gamma$ originates from space-group symmetry and remains intact within the studied range of UCS. Therefore, upon applying UCS $\geq$5\%, the four Weyl points [($\pm\frac{2\pi}{a}$(0.226), $\pm\frac{2\pi}{b}$(0.004), 0.000)] vanish.

To determine the nature of the Weyl point, we have calculated the 3D energy spectrum around the node point ($\frac{2\pi}{a}$(0.226), $\frac{2\pi}{b}$(0.004), 0.000). 
The corresponding results are shown in Fig. 7 of the SM~\cite{supplementry}. The 3D representation of the energy spectrum with respect to the wave vector \textbf{k} reveals that the Weyl point appears at the point of contact with a line-like constant energy surface. From the figure, it is seen that one of the energy bands is tilted and exactly aligns with a flat band without dispersion along the $k_z$=0 plane, serving as a characteristic feature of the type-III Weyl phase.
Along with this, we have also tested the robustness of the topological phases both in the absence and presence of SOC within the TB-mBJ potential. We find that, in the absence of SOC, a nodal-line phase emerges, whereas the Weyl phase remains intact in the presence of SOC under the TB-mBJ potential. Therefore, we conclude that the topological phases are robust against variations in exchange-correlation functionals.



To determine whether these degeneracies lead to topologically nontrivial states, we have further calculated the surface states and Fermi arcs along X-$\Gamma$-A high-symmetric direction. The corresponding results are shown in Figs. \ref{fig:surface_state} (a)-(f).
To calculate the surface states, one first obtains the tight-binding parameters for the bulk and surface atoms in order to study the non-trivial topological properties. 
The high-symmetry points (X, $\Gamma$ and A) of the 3D momentum space are projected onto the (001) surface. 
Therefore, the X, $\Gamma$ and A points are now denoted by $\bar{\mathrm{X}}$, $\bar{\Gamma}$ and $\bar{\mathrm{A}}$, respectively. This direction was chosen because linear touching is observed at the $\Gamma$ point. From the figures, it is evident that linear touchings occur between bands 1-$1^\prime$, 2-$2^\prime$, and 3-$3^\prime$ at energies of -185 meV, -104.5 meV, and 35.5 meV, respectively.
For the experimental detection of the topological character of the Weyl point, it is essential to analyze the surface spectrum that hosts Fermi arcs, as the presence of non-trivial Fermi arcs is a hallmark of Weyl semimetals. Thus, we have calculated the Fermi arc, as shown in Figs. \ref{fig:surface_state} (d)-(f), for bands 1-$1^\prime$ at an energy of -185 meV, bands 2-$2^\prime$ at -104.5 meV, and bands 3-$3^\prime$ at 35.5 meV. 
Since there is only one Weyl point located at $\Gamma$, it is expected that the Fermi arc originates and merges at the $\Gamma$ point, forming a closed loop. The results obtained in Fig. \ref{fig:surface_state} (d) are in accordance with this discussion.
From the figure, it can be seen that the non-trivial Fermi arc forms a closed loop.
The shape of the Fermi arcs for different combinations of bands changes as we move from band 1 to band $3^\prime$. However, the features of the Fermi arcs for bands 2-$2^\prime$ and 3-$3^\prime$ are the same as those of the Fermi arc for band 1-$1^\prime$. 
The existence of such Fermi arcs signifies a non-trivial topological phase in the system, and their presence is intrinsically linked to systems with a non-zero Chern number. 
It is also important to highlight that a doubly degenerate point is considered to possess a non-trivial topological feature e.g. Weyl semimetallic feature if the regions around degenerate point exhibits one or more defining features. First, the presence of linear dispersion in \textbf{k}-space around the node point is a necessary condition for a Weyl semimetal. Second, within the framework of a minimal model, the two touching bands carry opposite chiralities of +1 and -1. If $1^{\rm st}$ band carries +1 chirality then $2^{\rm nd}$ band will possess a chirality of -1. Moreover, chirality is calculated over a single band not over both the bands. If one tries to calculate the chirality over both the bands then result will be zero. The calculated Berry curvature corresponding to first band with +1 chirality will show a diverging behavior at node point whereas Berry curvature corresponding to second band with -1 chirality will show a converging behavior at node point. Third, the Nielsen-Ninomiya no-go theorem~\cite{nielsen1981no, nielsen1983adler} imposes the constraint that the total chirality of a material must vanish. Consequently, node points must come in pairs and chirality corresponding to each pair of node points of a single band must be $\pm$1. The bulk boundary condition provides the existence of non trivial surface states known as Fermi arcs. These arcs must connect the node points with opposite chirality. Based on the above properties, we have assigned chiralities of $\pm$1 to the touching point at $\Gamma$. Since we observe linear dispersion in the vicinity of $\Gamma$, the calculated chiralities corresponding to bands 1 \& $1^\prime$, 2 \& $2^\prime$, and 3 \& $3^\prime$ are found to be +1 \& -1, respectively. As the $\Gamma$ point is the center of the Brillouin zone, it should act both as a source and a sink of the Berry curvature; hence, one expects the corresponding Fermi arc to form a closed loop. Moreover, because the $\Gamma$ point lies at the Brillouin zone center, no equivalent point is generated upon applying space group symmetry, time-reversal symmetry, or both. In contrast, it is typically seen that other points in the Brillouin zone generate equivalent points when these symmetries are applied. Furthermore, the Nielsen-Ninomiya (no-go) theorem is not valid at the $\Gamma$ point ~\cite{Ma_2021}.
The calculated surface state indeed shows a loop originating from and terminating at the $\Gamma$ point.
A similar behavior is observed in this material at each compressive strain value (-5\%, -5.5\%, and -5.6\%). In addition, we have also discussed the topological surface states and Fermi arcs, which enclose 4 Weyl points ($\pm\frac{2\pi}{a}$(0.226), $\pm\frac{2\pi}{b}$(0.004), 0.000) (see Fig. 8 in the SM~\cite{supplementry}). 
Therefore, based on the above results, it is concluded that a Weyl point exists at the $\Gamma$ point. This theoretical prediction can potentially be verified through corresponding experimental studies. Furthermore, the predicted result highlights the potential of Cu$_2$SnS$_3$ for various practical applications. 
Along with this, we have also studied the transition from the type-II nodal line phase to the type-III Weyl phase using a model Hamiltonian, and found that the $\mu$ parameter mimics UCS (see Figs. 9-11 and the corresponding text in the SM~\cite{supplementry} for details).


{\it Conclusions and Outlook.|}
In summary, using \textit{ab-initio} calculations, we have investigated the evolution of multiple topological phases and their transitions as a function of uniaxial strain in orthorhombic Cu$_2$SnS$_3$. 
In the absence of SOC, the material shows a type-II nodal-ring and in the presence of SOC, it exhibit Weyl phase with seven Weyl points (three at $\Gamma$ and four at ($\pm\frac{2\pi}{a}$(0.226), $\pm\frac{2\pi}{b}$(0.004), 0.000) positions) along with nodal arcs.
Notably, type-II nodal-ring persists up to a UCS of $<5.5$\%. However, at $5.5$ ($5.6$)\% UCS, a type-III nodal-ring (Weyl point) emerges. 
The type-III nodal line phase realized in the present work is useful for searching materials with carriers residing in flat energy bands, leading to various emerging phenomena such as strong nonlinear optical responses and large magnetoresistance or magneto-elastoresistance.
In the presence of SOC, for 5\% $\leq$ UCS $<$ 5.6\%, four Weyl points at ($\pm\frac{2\pi}{a}$(0.226), $\pm\frac{2\pi}{b}$(0.004), 0.000) positions vanish, while nodal arcs remain intact throughout the studied range of UCS.
The presence of multiple topological phases in Cu$_2$SnS$_3$ and multiple topological phase transitions under uniaxial strain highlight its importance, making it an interesting material for future research.

\bibliography{MS}
\bibliographystyle{apsrev4-2}

\end{document}